\documentclass[letterpaper, 10pt, conference]{ieeeconf}
\usepackage[autostyle]{csquotes}
\usepackage{bm}
\usepackage{amssymb}
\usepackage{epsfig}
\usepackage{subfigure}
\usepackage{multicol, blindtext}
\usepackage{amsmath}
\usepackage{graphicx}
\usepackage{epstopdf}
\usepackage{float}
\usepackage{amsfonts}
\usepackage{color}
\usepackage{dblfloatfix}
\usepackage{nidanfloat}
\usepackage{multirow}%
\usepackage{fourier} 
\usepackage{array}
\usepackage{makecell}

\providecommand{\U}[1]{\protect\rule{.1in}{.1in}}

\newtheorem{theorem}{\rm\textbf{Theorem}}

\IEEEoverridecommandlockouts
\begin{document}

\title{{\LARGE \textbf{Optimal Control of Connected Automated Vehicles with Event-Triggered Control Barrier Functions }}
\thanks{This work was supported in part by NSF under grants ECCS-1931600,
DMS-1664644, CNS-1645681, by AFOSR under grant FA9550-19-1-0158,
by ARPA-E under grant DE-AR0001282 and by the NEXTCAR program
under grant DEAR0000796, and by the MathWorks and by NPRP grant
(12S-0228-190177) from the Qatar National Research Fund, a member of
the Qatar Foundation (the statements made herein are solely the responsibility
of the authors).
The authors are with the Division of Systems Engineering and Center for
Information and Systems Engineering, Boston University, Brookline, MA,
02446, USA and Electrical Engineering Department,Qatar University, Doha, Qatar \texttt{{\small \{esabouni, cgc,xiaowei\}@bu.edu, \{nader.meskin@qu.edu.qa\}}}}}

\author{Ehsan Sabouni, Christos G. Cassandras, Wei Xiao and Nader Meskin}
\maketitle

\begin{abstract}
We address the problem of controlling Connected and Automated Vehicles (CAVs) in conflict areas of a traffic network subject to hard safety constraints. It has been shown that such problems can be solved through a combination of tractable optimal control problem formulations and the use of Control Barrier Functions (CBFs) that guarantee the satisfaction of all constraints. These solutions can be reduced to a sequence of Quadratic Programs (QPs) which are efficiently solved on line over discrete time steps. However, the feasibility of each such QP cannot be guaranteed over every time step. To overcome this limitation, we develop an event-driven approach such that the next QP is triggered by properly defined events and
show that this approach can eliminate infeasible cases due to time-driven inter-sampling effects.
Simulation examples show how overall infeasibilities can be significantly reduced  with the proposed event-triggering scheme, while also reducing the need for communication among CAVs without compromising performance.

\end{abstract}

\thispagestyle{empty} \pagestyle{empty}


\section{INTRODUCTION}

The emergence of Connected and Automated Vehicles (CAVs) along with new traffic infrastructure technologies \cite{li2013survey} over the past decade have brought the promise of resolving long-lasting problems in transportation networks such as accidents, congestion, and unsustainable energy consumption along with environmental 
pollution \cite{f60c6852682d4629b75f458c50adbbf1},\cite{Schrank20152015UM}. Meeting this goal heavily depends on effective traffic management, specifically at the bottleneck points of a transportation network such as intersections, roundabouts, and merging roadways. 

To date, both centralized and decentralized methods have been proposed to tackle the control and coordination problem of CAVs in conflict areas; an overview of such methods may be found in \cite{7562449}. Platoon formation \cite{xu2019grouping} and reservation-based methods are among the centralized approaches, which are limited by the need for powerful central computation resources and are typically prone to disturbances and security threats.
In contrast, in decentralized methods each CAV is responsible for its own on-board computation with information from other vehicles limited to a set of neighbors \cite{7313484}. Constrained optimal control problems can then be formulated with objectives usually involving minimizing acceleration or maximizing passenger comfort (measured as the acceleration derivative or jerk), or jointly minimizing travel time through conflict areas and energy consumption. These problems can be analytically solved in some cases, e.g., for optimal merging \cite{XIAO2021109333} or crossing
a signal-free intersection \cite{Zhang2018}.
However, obtaining such solutions becomes computationally prohibitive for real-time applications when an optimal trajectory involves multiple constraints becoming active. Thus, Model Predictive Control (MPC) techniques are often adopted for real-time control execution and the handling of additional constraints \cite{cao2015cooperative} \cite{mukai2017model}. The use of Control Barrier Functions (CBFs) has been recently introduced as an alternative to MPC, largely due to their forward invariance property which provides crucial safety constraint guarantees \cite{Xiao2019}. 



An approach combining optimal control solutions with CBFs was recently presented in \cite{XIAO2021109592}. In this combined approach (termed OCBF), the solution of an \emph{unconstrained} optimal control problem is first derived and used as a reference control. Then, the resulting control reference trajectory is optimally tracked subject to
a set of CBF constraints which ensure the satisfaction of all constraints of the original optimal control problem. Finally, this optimal tracking problem is efficiently solved by discretizing time and solving a simple Quadratic Problem (QP) at each discrete time step over which the control input is held constant \cite{CBF_QP(2017)}. The use of CBFs in this approach exploits their forward invariance property to guarantee that all constraints they enforce are satisfied at all times if they are initially satisfied. In addition, CBFs are designed to impose \emph{linear} constraints on the control which is what enables the efficient solution of the tracking problem through a sequence of QPs. This approach can also be shown to provide additional flexibility in terms of using nonlinear vehicle dynamics (as long as they are affine in the control), complex objective functions, and tolerate process and measurement noise. However, the control update interval in the time discretization process must be sufficiently small in order to always guarantee that every QP is feasible. In practice, such feasibility can be often seen to be violated. One way to remedy this issue is to use an \emph{event-driven} scheme instead. The synthesis of event-triggered control and BFs or Lyapunov functions has been addressed before in \cite{ong2018event} with the goal of improving stability and
a unified event-driven scheme is proposed in \cite{taylor2020safety} with an Input-to-State barrier function to impose safety under an input disturbance. 

The contribution of this paper is to replace the \emph{time-driven} nature of the discretization that gives rise to QPs in the OCBF approach by an \emph{event-triggering} scheme, with the aim of achieving QP feasibility independent of a time step choice.
Given the system state at the start of a given QP instance, we follow the approach introduced in \cite{Xiao2021EventTriggeredSC} to define events associated with the states reaching a certain bound, at which point the next QP instance is triggered. We will show that this approach guarantees the forward invariance property of CBFs and eliminates infeasible cases due to time-driven inter-sampling effects (additional infeasibilities are still possible due to potentially conflicting constraints within a QP; this separate issue has been addressed in \cite{XIAO2022inf}). In addition, this approach can significantly reduce the number of QPs, thereby reducing the need for unnecessary communication among CAVs. Moreover, the unpredictability of event-triggering relative to a fixed time discretization approach can drastically reduce the potential for external malicious activity, hence improving security.

The paper is organized as follows. In Section II, we provide an overview of the decentralized constrained optimal control for CAVs in any conflict area setting, along with a brief review of CBFs to set the stage for the OCBF approach. We also review the time-driven approach for solving such optimal control problems, motivating the proposed event-driven approach. In Section III, this approach is presented, including the formulation and solution of QPs in an event-driven framework. In Section V, simulation results compare the time-driven and event-driven approaches in terms of infeasible cases and show how constraint violations can be reduced in the latter.


\section{Problem Formulation and Time-Driven Control Solutions}

\label{sec:problem}
\begin{figure}[H]
\vspace{-5mm}
\centering
$\hspace{-4mm}$\includegraphics[scale=0.8]{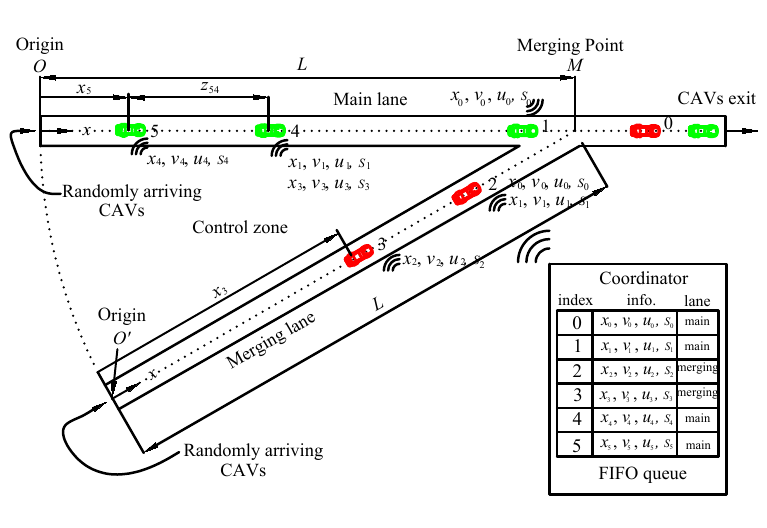} \caption{The merging problem}%
\label{fig:merging}%
\end{figure}

In this section, we review the setting for CAVs whose motion is cooperatively controlled at conflict areas of a traffic network. This includes merging roads, signal-free intersections, roundabouts, and highway segments where lane change maneuvers take place. We define a Control Zone (CZ) to be an area within which CAVs can communicate with each other or with a coordinator (e.g., a Road-Side Unit (RSU)) which is responsible for facilitating the exchange of information (but not control individual vehicles) within this CZ. As an example, Fig. \ref{fig:merging} shows a conflict area due to vehicles merging from two single-lane roads and there is a single Merging Point (MP) which vehicles must cross from either road \cite{XIAO2021109333}. More generally, the CZ may include a set of MPs that each CAV has to cross; for instance in a 4-way intersection with two lanes per direction there are 32 total MPs \cite{xu2020general}.

In such a setting, assuming all traffic consists of CAVs, a finite horizon constrained optimal control problem can be formulated aiming to determine trajectories that jointly minimize travel time and energy consumption through the CZ while also ensuring passenger comfort (by minimizing jerk or centrifugal forces) and guaranteeing safety constraints are always satisfied. 
Let $F(t)$ be the set of indices of all CAVs located in the
CZ at time $t$. A CAV enters the CZ at one of several origins (e.g., $O$ and $O'$ in Fig. \ref{fig:merging}) and leaves at one of possibly several exit points (e.g., $M$ in Fig. \ref{fig:merging}). 
The index $0$ is used to denote a CAV that has just left the CZ. Let $N(t)$ be the
cardinality of $F(t)$. Thus, if a CAV arrives at time $t$, it is assigned the
index $N(t)+1$. All CAV indices in $F(t)$ decrease by one when a CAV passes over
the MP and the vehicle whose index is $-1$ is dropped.

The vehicle dynamics for each CAV $i\in F(t)$ along the lane to which it
belongs in a given CZ are assumed to be of the form
\begin{equation} \label{VehicleDynamics}
\left[
\begin{array}
[c]{c}%
\dot{x}_{i}(t)\\
\dot{v}_{i}(t)
\end{array}
\right]  =\left[
\begin{array}
[c]{c}%
v_{i}(t)\\
u_{i}(t)
\end{array}
\right],  
\end{equation}
where $x_{i}(t)$ denotes the distance from the origin at which CAV $i$ arrives, $v_{i}(t)$ denotes the velocity, and $u_{i}(t)$ denotes the control input (acceleration).
There are two objectives for each CAV, as detailed next.\\
\noindent {\bf Objective 1} (Minimize travel time): Let $t_{i}^{0}$ and $t_{i}^{f}$
denote the time that CAV $i\in F(t)$ arrives at its origin
and leaves the CZ at its exit point, respectively. We wish to minimize the travel time
$t_{i}^{f}-t_{i}^{0}$ for CAV $i$.\\
{\bf Objective 2} (Minimize energy consumption): We also wish to minimize
the energy consumption for each CAV $i$:
\begin{equation}
J_{i}(u_{i}(t),t_{i}^{f})=\int_{t_{i}^{0}}^{t_{i}^{f}}\mathcal{C}_i(u_{i}(t))dt,
\end{equation}
where $\mathcal{C}_i(\cdot)$ is a strictly increasing function of its argument.

A comfort objective may also be included (e.g., when the CZ includes curved road segments subject to centrifugal forces \cite{9682916}, but we omit it here for simplicity). We consider next the following constraints.\\
{\bf Constraint 1} (Safety constraints): Let $i_{p}$ denote the index of
the CAV which physically immediately precedes $i$ in the CZ (if one is
present). We require that the distance $z_{i,i_{p}}(t):=x_{i_{p}}(t)-x_{i}(t)$
be constrained by:
\begin{equation}
z_{i,i_{p}}(t)\geq\varphi v_{i}(t)+\delta,\text{ \ }\forall t\in\lbrack
t_{i}^{0},t_{i}^{f}], \label{Safety}%
\end{equation}
where $\varphi$ denotes the reaction time (as a rule, $\varphi=1.8s$ is used,
e.g., \cite{Vogel2003}) and $\delta$ is a given minimum safe distance.
 If we define $z_{i,i_{p}}$ to be the distance from
the center of CAV $i$ to the center of CAV $i_{p}$, then $\delta$ depends on the length of these two CAVs (generally dependent on
$i$ and $i_{p}$ but taken to be a constant over all CAVs for simplicity).\\
{\bf Constraint 2} (Safe merging): Whenever a CAV crosses a MP (including one that varies dynamically if a CAV changes lanes as in 
\cite{xu2020general}, a lateral collision is possible and there must be adequate safe space for the CAV at this MP to avoid such collision, i.e.,
\begin{equation}
\label{SafeMerging}
z_{i,j}(t_{i}^{m})\geq\varphi v_{i}(t_{i}^{m})+\delta,
\end{equation}
where $j$ is the index of the CAV that may collide with CAV $i$ at merging point $m=\lbrace 1,...,n_i \rbrace$ where $n_i$ is the total number of MPs that CAV $i$ passes in the CZ. The determination of CAV $j$ depends on the policy adopted for sequencing CAVs through the CZ, such as First-In-First-Out (FIFO) based on the arrival times of CAVs, the Dynamic Resequencing (DR) policy presented in \cite{Zhang2018} or any other desired policy. It is worth noting that this constraint only applies at a certain time $t_{i}^{m}$ which obviously depends on how the CAVs are controlled. As an example, in Fig. \ref{fig:merging} under FIFO, we have $j=i-1$ and $t_i^m=t_i^f$ since the MP defines the exit from the CZ.\\
{\bf Constraint 3} (Vehicle limitations): Finally, there are constraints
on the speed and acceleration for each $i\in F(t)$:
\begin{equation}
\begin{aligned} v_{imin} \leq v_i(t)\leq v_{imax}, \forall t\in[t_i^0,t_i^f]\end{aligned} \label{VehicleConstraints1}%
\end{equation}
\begin{equation}
\begin{aligned} u_{i{min}}\leq u_i(t)\leq u_{i{max}}, \forall t\in[t_i^0,t_i^f],\end{aligned} \label{VehicleConstraints2}%
\end{equation}
where $v_{imax}> 0$ and $v_{imin} \geq 0$ denote the maximum and minimum speed allowed
in the CZ for CAV $i$, $u_{i{min}}<0$ and $u_{i{max}}>0$ denote the minimum and maximum
control for CAV $i$, respectively.\\
\textbf{Optimal Control Problem formulation.} Our goal is to determine a control law achieving objectives 1-2 subject to constraints 1-3 for each $i \in F(t)$ governed by the dynamics (\ref{VehicleDynamics}). Choosing $\mathcal{C}_i(u_i(t))=\frac{1}{2}u_i^2(t)$ and normalizing travel time and $\frac{1}{2}u_{i}^{2}(t)$, we use the weight $\alpha\in[0,1]$ to construct
a convex combination as follows:
\begin{equation}\label{eqn:energyobja_m}
\begin{aligned}\min_{u_{i}(t),t_i^f} J_i(u_i(t),t_i^f)= \int_{t_i^0}^{t_i^f}\left(\alpha + \frac{(1-\alpha)\frac{1}{2}u_i^2(t)}{\frac{1}{2}\max \{u_{imax}^2, u_{imin}^2\}}\right)dt \end{aligned}.
\end{equation}
 Letting $\beta:=\frac{\alpha\max\{u_{imax}^{2},u_{imin}^{2}\}}{2(1-\alpha)}$, we obtain a simplified form: 
\begin{equation}\label{eqn:energyobja}
\min_{u_{i}(t),t_i^f}J_{i}(u_{i}(t),t_i^f):=\beta(t_{i}^{f}-t_{i}^{0})+\int_{t_{i}^{0}%
}^{t_{i}^{f}}\frac{1}{2}u_{i}^{2}(t)dt,
\end{equation}
where $\beta\geq0$ denotes a weight factor that can be adjusted to penalize
travel time relative to the energy cost. Note that the solution is \emph{decentralized} in the sense that CAV $i$ requires information only from CAVs $i_p$ and $j$ required in (\ref{Safety}) and (\ref{SafeMerging}).

Problem (\ref{eqn:energyobja}) subject to (\ref{VehicleDynamics}), (\ref{Safety}), (\ref{SafeMerging}), (\ref{VehicleConstraints1})
and (\ref{VehicleConstraints2}) can be analytically solved in some cases, e.g., the merging problem in Fig. \ref{fig:merging}
\cite{XIAO2021109333}
and a signal-free intersection 
\cite{Zhang2018}.
However, obtaining solutions for real-time applications becomes prohibitive when an optimal trajectory involves multiple constraints becoming active. This has motivated an approach which combines a solution of the unconstrained problem (\ref{eqn:energyobja}), which can be obtained very fast, with the use of Control Barrier Functions (CBFs) which provide guarantees that (\ref{Safety}), (\ref{SafeMerging}), (\ref{VehicleConstraints1}) and (\ref{VehicleConstraints2}) are always satisfied through constraints that are linear in the control, thus rendering solutions to this alternative problem obtainable by solving a sequence of computationally efficient QPs. This approach is termed Optimal Control with Control Barrier Functions (OCBF) \cite{XIAO2021109592}.

\textbf{The OCBF approach.} The OCBF approach consists of three steps: (i) the solution of the \emph{unconstrained} optimal control problem (\ref{eqn:energyobja}) is used as a reference control, (ii) the resulting control reference trajectory is optimally tracked subject to the constraint \eqref{VehicleConstraints2}, as well as a set of CBF constraints enforcing (\ref{Safety}), (\ref{SafeMerging}) and \eqref{VehicleConstraints1}. (iii) This optimal tracking problem is efficiently solved by discretizing time and solving a simple QP at each discrete time step. The significance of CBFs in this approach is twofold: first, their forward invariance property \cite{XIAO2021109592} guarantees that all constraints they enforce are satisfied at all times if they are initially satisfied; second, CBFs impose \emph{linear} constraints on the control which is what enables the efficient solution of the tracking problem through the sequence of QPs in (iii) above.

The reference control in step (i) above is denoted by $u_{iref}(t)$. The unconstrained solution to (\ref{eqn:energyobja}) is denoted by $u_i^*(t)$, thus we usually set $u_{iref}(t)=u_i^*(t)$. However, $u_{iref}(t)$ may be any desired control trajectory and, in general, we use $u_{iref}(t)=h(u_i^*(t),x_i^*(t), \textbf{x}_i(t))$ where 
$\textbf{x}_i(t)\equiv(x_i(t),v_i(t)), \textbf{x}_i \in \textbf{X}$ ($\mathbf{X} \subset \mathbb{R}^2$ is the state space). Thus, in addition to the unconstrained optimal control and position $u_i^*(t),x_i^*(t)$, observations of the actual CAV state $\textbf{x}_i(t)$ provide direct feedback as well. 

To derive the CBFs that ensure the constraints (\ref{Safety}), (\ref{SafeMerging}), and (\ref{VehicleConstraints1}) are always satisfied, we use the vehicle dynamics (\ref{VehicleDynamics}) to define $f(\textbf{x}_i(t))=[v_i(t),0]^T$ and $g(\textbf{x}_i(t))=[0,1]^T$. Each of these constraints can be easily written in the form of $b_q(\textbf{x}(t)) \geq 0$, $q \in \lbrace  1,...,n \rbrace$ where $n$ stands for the number of constraints and $\mathbf{x}(t)=[\mathbf{x}_1(t),\mathbf{x}_2(t),...,\mathbf{x}_{N(t)}(t)]$. The CBF method (details provided in \cite{XIAO2021109592}) maps a constraint $b_q(\textbf{x}(t)) \geq 0$ onto a new constraint which is linear in the control input $u_i(t)$ and takes the general form 
\begin{equation} \label{CBF general constraint}
L_fb_q(\textbf{x}(t))+L_gb_q(\textbf{x}(t))u_i(t)+\gamma( b_q(\textbf{x}(t))) \geq 0,
\end{equation}
where $L_f,L_g$ denote the Lie derivatives of $b_q(\textbf{x}(t))$ along $f$ and $g$ respectively and $\gamma(\cdot)$ stands for any class-$\mathcal{K}$ function \cite{XIAO2021109592}. It has been established 
\cite{XIAO2021109592}
that satisfaction of (\ref{CBF general constraint}) implies the satisfaction of the original problem constraint $b_q(\textbf{x}(t)) \geq 0$ because of the forward invariance property. It is worth observing that the newly obtained constraints are sufficient conditions for the original problem constraints, therefore, potentially conservative. We also note that if the relative degree of the function $b_q(\textbf{x}(t))$ (i.e., the number of times it needs to be
differentiated along its dynamics until the control $u$ explicitly shows in the corresponding derivative)
in (\ref{CBF general constraint}) is greater than one, then we need to make use of High Order CBFs (HOCBFs) as detailed in \cite{XIAO2021109592}.
This situation does not arise in the problems considered in this paper.

We now apply (\ref{CBF general constraint}) to obtain the CBF constraint associated with the safety constraint (\ref{Safety}). By setting $b_1(\textbf{x}_i(t),\textbf{x}_{i_p}(t))=z_{i,i_{p}}(t)-\varphi v_{i}(t)-\delta=x_{i_p}(t)-x_i(t)-\varphi v_i(t)-\delta$ and since $b_1(\textbf{x}_i(t),\textbf{x}_{i_p}(t))$ is differentiable,
the CBF constraint for (\ref{Safety}) is
\begin{equation}\label{CBF1}\small
\underbrace{v_{i_p}(t)-v_i(t)}_{L_fb_1(\textbf{x}_i(t),\textbf{x}_{i_p}(t))}+\underbrace{-\varphi}_{L_gb_1(\textbf{x}_i(t))} u_i(t)+\underbrace{k_1(z_{i,i_p}(t)-\varphi v_i(t))}_{\gamma_1(b_1(\textbf{x}_i(t),\textbf{x}_{i_p}(t)))} \geq 0,
\end{equation}
where the class-$\mathcal{K}$ function $\gamma(x)=k_1x$ is chosen here to be linear.

Deriving the CBF constraint for the safe merging constraint (\ref{SafeMerging}) poses a technical challenge due to the fact that it only applies at a certain time $t_i^{m}$, whereas a CBF is required to be in a continuously differentiable form. To tackle this problem. we apply a technique used in \cite{XIAO2021109592} to convert (\ref{SafeMerging}) to a continuous differentiable form as follows:
\begin{equation}
z_{i,j}(t_{i}^{m})-\Phi(x_i(t)) v_{i}(t_{i}^{m})-\delta \geq 0, \   {\ }\forall t\in[t_i^0,t_i^{m}],
\end{equation}
where $\Phi : \mathbb{R} \rightarrow \mathbb{R}$ may be any continuously differentiable function as long as it is strictly increasing and satisfies the boundary conditions $\Phi(x_i(t_i^0))=0$ and $\Phi(x_i(t_i^{m}))=\varphi$. In this case, a linear function can satisfy both conditions:
\begin{equation}
    \Phi(x_i(t))=\varphi  \frac{x_i(t)}{L}
\end{equation}
where $L$ is the length of road traveled by the CAV from its entry to the CZ to the MP of interest in (\ref{SafeMerging}).
Then, proceeding as in the derivation of (\ref{CBF1}), we obtain:
\begin{align}\small \label{CBF2}
&\underbrace{v_{j}(t)-v_i(t)-\frac{\varphi}{L}v_i^2(t)}_{L_fb_2(\textbf{x}_i(t),\textbf{x}_j(t))}+\underbrace{-\varphi \frac{x_i(t)}{L}}_{L_gb_2(\textbf{x}_i(t))}u_i(t)+\nonumber \\ &\underbrace{k_2(z_{i,j}(t)-\varphi  \frac{x_i(t)}{L} v_i(t)}_{\gamma_2(b_2(\textbf{x}_i(t),\textbf{x}_j(t)))} \geq 0.
\end{align}

The speed constraints in \eqref{VehicleConstraints1} are also easily transformed into CBF constraints using (\ref{CBF general constraint}) by defining $b_3(\textbf{x}_i(t))=v_{imax}-v_i(t)$ and $b_4(\textbf{x}_i(t))=v_i(t)-v_{imin}$. This yields:
\begin{align} \label{CBF3-4}
\underbrace{-1}_{L_gb_3(\textbf{x}_i(t))}u_i(t)+\underbrace{k_3(v_{imax}-v_i(t))}_{\gamma_3(b_3(\textbf{x}_i(t)))} \geq 0 \nonumber, \\
\underbrace{1}_{L_gb_4(\textbf{x}_i(t))}u_i(t)+\underbrace{k_4(v_i(t)-v_{imin})}_{\gamma_4(b_4(\textbf{x}_i(t)))} \geq 0,
\end{align}
for the maximum and minimum velocity constraints, respectively.

As a last step in the OCBF approach, we can exploit the versatility of the CBF method by using a Control Lyapunov Function (CLF) to track specific state variable in the reference trajectory if desired. A CLF $V(\textbf{x}_i(t))$ is similar to a CBF (see \cite{XIAO2021109592}). In our problem, letting $V(\textbf{x}_i(t))=(v_i(t)-v_{i{ref}}(t))^2$ we can express the CLF constraint associated with tracking the CAV speed to a desired value $v_{i{ref}}(t)$ (if one is provided) as follows:
\begin{equation}\label{CLF}
L_fV(\textbf{x}_i(t))+L_gV(\textbf{x}_i(t))u_i(t)+\epsilon V(\textbf{x}_i(t))\leq e_i(t),
\end{equation}
where $\epsilon >0 $ and $e_i(t)$ makes this a soft constraint.

Now that all the original problem constraints have been transformed into CBF constraints, we can formulate the OCBF problem as follows:
\begin{equation}\label{QP-OCBF}\small
\min_{u_i(t),e_i(t)}J_i(u_i(t),e_i(t)):=\int_{t_i^0}^{t_i^f}\big[\frac{1}{2}(u_i(t)-u_{iref})^2+\lambda e^2_i(t)\big]dt
\end{equation}
subject to vehicle dynamics (\ref{VehicleDynamics}), the CBF constraints (\ref{CBF1}), (\ref{CBF2}),\eqref{CBF3-4}, and \eqref{VehicleConstraints2} and CLF constraint (\ref{CLF}).

A common way to solve this dynamic optimization problem is to discretize $[t_i^0,t_i^f]$ into intervals $[t_i^0,t_i^0+\Delta],...,[t_i^0+k\Delta,t_i^0+(k+1)\Delta],...$ with equal length $\Delta$ and solving (\ref{QP-OCBF}) over each time interval. The decision variables $u_{i,k}=u_i(t_{i,k})$ and $e_{i,k}=e_i(t_{i,k})$ are assumed to be constant on each interval and can be easily calculated at time $t_{i,k}=t_i^0+k\Delta$ through solving a QP at each time step:
\begin{align} \label{QP}
\min_{u_{i,k},e_{i,k}}&[ \frac{1}{2}(u_{i,k}-u_{iref}(t_{i,k}))^2+\lambda e_{i,k}^{2}]
\end{align} 
subject to the CBF constraints (\ref{CBF1}), (\ref{CBF2}), \eqref{CBF3-4} and control input bounds \eqref{VehicleConstraints2} and CLF constraint (\ref{CLF}) where all constraints are linear in the decision variables. We refer to this as the \emph{time-driven} approach, which is fast and can be readily used in real time.

The main problem with this approach is that any one QP may become infeasible because the decision variable $u_{i,k}$ is held constant over a given time period $\Delta$. Since this is externally defined, there is no guarantee that it is small enough to ensure the forward invariance property of a CBF, thereby also failing to ensure the satisfaction of the safety constraints. In other words, in this time-driven approach, there is a critical (and often restrictive) assumption that the control update rate is high enough to avoid such a problem. There are several additional issues worth mentioning: (i) imposing a high update rate makes the solution of multiple QPs inefficient since it increases the computational burden, (ii) using a common update rate across all CAVs renders their synchronization difficult, and (iii) the predictability of a time-driven communication mechanism across CAVs makes the whole system susceptible to malicious attacks.  
Our proposed resolution of this problem is to use an \emph{event-driven} approach, as described next.\\
\textbf{}

\section{Event-Triggered Control}
\label{sec:event-triggered}
There are several possible event-driven mechanisms one can adopt to invoke the solution of the QPs in (\ref{QP}) subject to the CBF constraints (\ref{CBF1}), (\ref{CBF2}) and \eqref{CBF3-4} along with control input bounds \eqref{VehicleConstraints2}. One idea is to create a \emph{self-triggering} framework with a minimum inter-event time guarantee by 
predicting the first time instant that any of the CBF constraints in the QP problem (\ref{QP}) is violated at $t_{i,k}$ and select that as the next
time instant $t_{i,k+1}$ when CAV $i$ communicates with the coordinator and updates the control;
this approach is followed in [XXX]. 
In this paper, we adopt a different \emph{event-triggering} scheme introduced in \cite{Xiao2021EventTriggeredSC} such that we only need to solve a QP (with its associated CBF constraints) when one of three possible events (as defined next) is detected. We will show that this provides a guarantee for the satisfaction of the safety constraints which cannot be offered by the time-driven approach described earlier. The key idea is to ensure that the safety constraints are satisfied while the state remains within some bounds and define events which coincide with the state reaching these bounds, at which point the next instance of the QP in (\ref{QP}) is triggered.

Let $t_{i,k}$, $k=1,2,...$, be the time instants when the QP in (\ref{QP}) is solved. Our goal is to guarantee that the state trajectory does not violate any safety constraints within any time interval $(t_{i,k},t_{i,k+1}]$ where $t_{i,k+1}$ is the next time instant when the QP is solved. Define a subset of the state space of CAV $i$ at time $t_{i,k}$ such that:
\begin{equation} \label{bound}
\textbf{x}_i(t_{i,k})-\textbf{s}_i \leq \textbf{x}_i(t) \leq \textbf{x}_i(t_{i,k})+\textbf{s}_i,
\end{equation}
where $\textbf{s}_i =\left[s_{i_x} \ \ s_{i_v} \right]^T \in \mathbb{R}_{>0}^2$ is a parameter vector whose choice will be discussed later. Intuitively, this choice reflects a trade-off between computational efficiency (when the values are large and there are fewer instances of QPs to be solved) and conservativeness (when the values are small). We denote the set of states of CAV $i$ that satisfy \eqref{bound} at time $t_{i,k}$ by 
\begin{equation} \label{event bound}
S_i(t_{i,k}) = \lbrace\textbf{y}_i \in \textbf{X}: \textbf{x}_i(t_{i,k})-\textbf{s}_i \leq \textbf{y}_i \leq \textbf{x}_i(t_{i,k})+\textbf{s}_i\rbrace.
\end{equation} 
In addition, let $C_{i,1}$ be the feasible set of our original constraints \eqref{Safety}, \eqref{SafeMerging} and \eqref{VehicleConstraints1} defined as
\begin{equation}
    C_{i,1}:=\lbrace \mathbf{x}_i\in \mathbf{X}:b_q(\mathbf{x})\geq 0, \ q \in \lbrace 1,2,3,4 \rbrace \rbrace,
\end{equation}
Next, we seek a bound and a control law that satisfies the safety constraints within this bound. This can be accomplished by considering the minimum value of each component in \eqref{CBF general constraint} for every $q \in \lbrace 1,2,3,4 \rbrace $  as shown next.

Let us start with the first term in \eqref{CBF general constraint}, $L_fb_q(\textbf{x}(t))$, rewritten as $L_fb_q(\textbf{y}_i(t),\textbf{y}_r(t))$ with $\mathbf{y}_i(t)$ as in (\ref{event bound}) and and $r$ stands for \enquote{relevant} CAVs affecting the constraint of $i$
(e.g., CAVs $i_p$ and $j$ in Fig. \ref{fig:merging}).
Let $b_{q,f_i,min}(t_{i,k})$ be the minimum possible value of the term $L_fb_q(\textbf{x}(t))$ 
over the time interval $(t_{i,k},t_{i,k+1}]$ for each $q= \lbrace 1,2,3,4 \rbrace $ over the set $\Bar{S_i}({t_{i,k}}) \cap \Bar{S_r}({t_{i,k}})$:
\begin{equation}\label{minfi}
b_{q,f_i,min}(t_{i,k})=\displaystyle\min_{\textbf{y}_i \in \Bar{S}_i({t_{i,k}}) \atop \textbf{y}_r \in \Bar{S}_r({t_{i,k}})}L_fb_q(\textbf{y}_i(t),\textbf{y}_r(t)),
\end{equation} 
where $\Bar{S}_i({t_{i,k}})$ is defined as follows:
\begin{equation}
    \Bar{S}_i({t_{i,k}}):=\lbrace\mathbf{y}_i \in C_{i,1} \cap S_i(t_{i,k}) \rbrace
\end{equation}

Similarly, we can define the minimum value of the third term in \eqref{CBF general constraint}:
\begin{equation}\label{mingammai}
b_{\gamma_q,min}(t_{i,k})=\displaystyle\min_{\textbf{y}_i \in \Bar{S}_i({t_{i,k}}) \atop \textbf{y}_r \in \Bar{S}_r({t_{i,k}})} \gamma_q(\textbf{y}_i(t),\textbf{y}_r(t)).
\end{equation} 

For the second term in \eqref{CBF general constraint}, note that $L_gb_q(\mathbf{x}_i)$ is a constant for $ q=\{1,3,4\} $, therefore there is no need for any minimization. However, $L_gb_2(\mathbf{x}_i)$ in (\ref{CBF2}) is state dependent and needs to be considered for the minimization. Since $x_i(t) \ge 0$, note that $L_gb_2(\mathbf{x}_i)=-\varphi \frac{x_i(t)}{L}$ is always negative, therefore, we can determine the limit value $b_{2,g_i,min}(t_{i,k}) \in \mathbb{R}, $ as follows:
\begin{eqnarray}\label{mingi} \small
b_{2,g_i,min}(t_{i,k})=\begin{cases}
\displaystyle\min_{\textbf{y}_i \in \Bar{S}_i({t_{i,k}}) \atop \textbf{y}_r \in \Bar{S}_r({t_{i,k}})}L_gb_2(\textbf{x}_i(t)), \  \textnormal{if}\  u_{i,k} \geq 0\\
\\
\displaystyle\max_{\textbf{y}_i \in \Bar{S}_i({t_{i,k}}) \atop \textbf{y}_r \in \Bar{S}_r({t_{i,k}})}L_gb_2(\textbf{x}_i(t)), \ \ \  \textnormal{otherwise},
\end{cases}
\end{eqnarray}
where the sign of $u_{i,k},  \ i \in F(t_{i,k})$ can be determined by simply solving the CBF-based QP  \eqref{QP} at time $t_{i,k}$.

Thus, the condition that can guarantee the satisfaction of \eqref{CBF1}, \eqref{CBF2} and \eqref{CBF3-4} in the time interval $\left(t_{i,k},t_{i,k+1}\right]$ is given by
\begin{equation} \label{minCBF}
b_{q,f_i,min}(t_{i,k})+b_{q,g_i,min}(t_{i,k})u_{i,k}+b_{\gamma_q,min}(t_{i,k})\geq 0,
\end{equation}
for $q=\lbrace1,2,3,4\rbrace$. In order to apply this condition to the QP \eqref{QP}, we just replace \eqref{CBF general constraint} by \eqref{minCBF} as follows:
\begin{align} \label{eq:QPtk}
\min_{u_{i,k},e_{i,k}}&[ \frac{1}{2}(u_{i,k}-u_{ref}(t_{i,k}))^2+\lambda e_{i,k}^{2}]\nonumber\\
 &\textnormal{s.t.} \ \  \eqref{CLF},\eqref{minCBF},\eqref{VehicleConstraints2}
\end{align} 
It is important to note that each instance of the QP \eqref{eq:QPtk} is now triggered by one of the following three events where $k =1,2,\ldots$ is an event (rather than time step) counter:\\
\textbf{Event 1:} the state of CAV $i$ reaches the boundary of $S_i(t_{i,k-1})$.
\\
\textbf{Event 2:} the state of CAV $i_p$ reaches the boundary of $S_{i_p}(t_{i,k-1})$ (if $i_p$ exists for CAV $i$).
\\
\textbf{Event 3:} the state of CAV $j$ reaches the boundary of $S_{j}(t_{i,k-1})$ where $j$ is the index of the CAV that may collide with $i$ in (\ref{SafeMerging}), e.g., 
$j=i-1 \neq i_p$ in the merging problem case, if such a CAV exists.

As a result, $t_{i,k},k=1,2,...$ is unknown in advance but can be determined by CAV $i$ through:
\begin{align} \label{events}
t_{i,k}=\min \Big\{ t>t_{i,k-1}:\vert\textbf{x}_i(t)-\textbf{x}_i(t_{i,k-1})\vert=\textbf{s}_i \\ \nonumber
\text{or} \ \ \vert\textbf{x}_{i_p}(t)-\textbf{x}_{i_p}(t_{i,k-1})\vert=\textbf{s}_{i_p} \\ \nonumber 
\text{or} \ \ \vert\textbf{x}_{j}(t)-\textbf{x}_{j}(t_{i,k-1})\vert=\textbf{s}_{j}\Big\},
\end{align}
where $t_{i,1}=0$. The events can be detected through the dynamics in \eqref{VehicleDynamics} or from on-board state measurements if available, along with state information from relevant other CAVs (e.g., CAVs $i_p$ and $j$ in Fig. \ref{fig:merging}) through the coordinator. Finally, note that because of the Lipschitz continuity of the dynamics in \eqref{VehicleDynamics} and the fact that the control is constant within an inter-event interval, Zeno behavior does not occur in this framework.

The selection of the parameters $\mathbf{s}_i$ captures the trade-off between computational cost and conservativeness: the larger the value of each component of $\mathbf{s}_i$ is, the smaller the number of events that trigger instances of the QPs becomes, thus reducing the total computational cost. At the same time, the control law must satisfy the safety constraints over a longer time interval as we take the minimum values in \eqref{minfi}-\eqref{mingi}, hence rendering the approach more conservative.

The following theorem formalizes our analysis by showing that if new constraints of the general form \eqref{minCBF} hold, then our original CBF constraints \eqref{CBF1}, \eqref{CBF2} and \eqref{CBF3-4} also hold. The proof follows the same lines as that of a more general theorem in \cite{Xiao2021EventTriggeredSC} and, therefore, is omitted.

\begin{theorem}\label{as:1} Given a CBF $b_q(\mathbf{x(t)})$ with relative degree one, let $t_{i,k+1}$, $k=1,2,\ldots$ be determined by \eqref{events} with $t_{i,1}=0$ and $b_{q,f_i,min}(t_{i,k})$, $b_{\gamma_q,min}(t_{i,k})$, $b_{q,g_i,min}(t_{i,k})$  for $q=\{1,2,3,4\}$ obtained through \eqref{minfi}, \eqref{mingammai}, and \eqref{mingi}. Then, any control input $u_{i,k}$ that satisfies \eqref{minCBF} for all $q \in \lbrace 1,2,3,4 \rbrace$  within the time interval $[t_{i,k},t_{i,k+1})$ renders the set $C_{i,1}$ forward invariant for the dynamic system defined in (\ref{VehicleDynamics}).
\end{theorem}

\textbf{Remark}1:
Expressing \eqref{minCBF} in terms of the minimum value of each component separately may become overly conservative if each minimum value corresponds to different points in the decision variable space. Therefore, an alternative approach is to calculate the minimum value of the whole term. 

{\bf Communication Protocol}.
As mentioned earlier, a coordinator is responsible for exchanging information among CAVs. In an event-driven scheme, 
frequent communication is generally not needed, since it occurs only when an event is triggered. Each CAV is responsible for checking its own state to detect any violation in its state bounds. When such an event occurs, the CAV updates its control input by re-solving the QP and inform the coordinator with its newly obtained state (i.e., velocity and position). It then becomes the responsibility of the coordinator to provide this information to the relevant CAVs (i.e., those that might be affected). Finally, the notified CAVs decide whether they need to re-solve their QP or maintain their control input until the next triggering event. Note that a triggered event due to an update in CAV $i$'s state can affect only CAVs $l>i$ and the ensuing \enquote{event chain}
will (at a maximum) stop once $l$ reaches the last CAV $l=N(t)$. 
As an example, suppose CAV $1$ in Fig. \ref{fig:merging}  solves its QP as a result of an event. Then, the newly obtained state is sent to the coordinator and forwarded to CAVs $2$ and $4$. Next, this new state will be checked separately by CAVs $2$ and $4$ for a potential bound violation. As a result, they either need to re-solve their QP, hence following the same procedure as CAV $1$ and sending their updated state to the coordinator or they carry on with their current control input until the next event. \\

\section{SIMULATION RESULTS}
\label{sec:simulation}

All algorithms in this section have been implemented using \textsc{MATLAB}.
We used \textsc{quadprog} for solving QPs of the form
\eqref{QP} and \eqref{eq:QPtk}, \textsc{lingprog} for solving the linear programming in \eqref{minfi}, \eqref{mingammai} and \eqref{mingi}, \textsc{fmincon} for a nonlinear optimization problem arising when \eqref{minfi} and \eqref{mingammai} become nonlinear, and \textsc{ode45} to integrate the vehicle dynamics.

We have considered the merging problem shown in Fig. \ref{fig:merging} where CAVs are simulated according to Poisson arrival processes with an arrival rate which is fixed for the purpose of making comparisons between the time-driven approach and the event-driven scheme (over different bound values in \eqref{events}). The initial speed $v_{i}(t_{i}^{0})$ is also randomly generated with a uniform distribution over $[15 \textnormal{m/s}, 20\textnormal{m/s}]$ at the origins $O$ and $O^{\prime}$, respectively. The
parameters for \eqref{QP-OCBF} and \eqref{eq:QPtk}
 are: $L = 400\textnormal{m}, \varphi = 1.8\textnormal{s}, \delta = 0\textnormal{m}, u_{max} = 4.905 \textnormal{m/s}^2, u_{min} = -5.886\textnormal{m/s}^2, v_{max} = 30\textnormal{m/s}, v_{min} = 0\textnormal{m/s},  k_1=k_2=k_3=k_4=1,  \lambda= 10$. The sensor sampling rate is $20$Hz, sufficiently high to avoid missing any triggering event (in general, a proper sampling rate can always be calculated given the CAV specifications, i.e., bounds on velocity and acceleration).
 The control update period for the time-driven control is $\Delta t=0.05$s. We let the bounds $S=[s_x,s_v]$ be the same for the all CAVs in the network and vary them between the values of $\lbrace[0.5,1.5],[0.5,2],[0.5,2.5]\rbrace$ to allow a comprehensive comparison. 
 
In our simulations, we included the computation of a more realistic energy consumption model \cite{kamal2012model} to supplement the simple surrogate $L_2$-norm ($u^2$) model in our analysis:
$
f_v(t)=f_{cruise}(t)+f_{accel}(t), 
f_{cruise}(t)= \omega_0+\omega_1v_i(t)+\omega_2v^2_i(t)+\omega_3v^3_i(t), 
f_{accel}(t)=(r_0+r_1v_i(t)+r_2v^2_i(t))u_i(t)
$
where we used typical values for parameters $\omega_1,\omega_2,\omega_3,r_0,r_1$ and, $r_2$ as reported in \cite{kamal2012model}.

Table I summarizes our results from 20 separate simulations corresponding to three different methods under the same conditions with different values for the relative weight of energy vs time as shown in the table: the time-driven method, the event-triggering scheme developed in this paper, and a self-triggering scheme described in a companion paper [XXX].
We observe that by using the event-triggering and self-triggering approaches we are able to significantly reduce the number of infeasible QP cases (up to $95\%$) compared to the time-driven approach. At the same time, the overall number of instances when a QP needs to be solved has also decreased up to $68\%$ and $80\%$ in the event-triggered and self-triggered approach respectively. 
Note that the large majority of infeasibilities is due to holding acceleration constant over an inappropriate sampling time, which can invalidate the forward invariance property of CBFs over the entire time interval. These infeasible cases were eliminated by the event-triggering and self-triggering schemes. However, another source of infeasibility is due to  conflicts that may arise between the CBF constraints and the control bounds in a QP. This cannot be remedied through the proposed event-triggered or self triggered QPs; it can, however, be dealt with by the introduction of a sufficient condition that guarantees no such conflict, as described in \cite{XIAO2022inf}. 

In Table I, we can also observe some loss of performance in the event-triggering method as the values of the bound parameters increase, hence increasing conservativeness. On the other hand, this decreases the computational load expressed in terms of the number of QPs that are solved, illustrating the trade-off discussed in the previous section. Likewise, the performance of the self-triggered scheme with minimum and maximum inter-event times $T_d$, $T_{max}$, respectively, is shown in the table (details on the self-triggered scheme and its relative advantages and disadvantages can be found in [XXX]).
There is also an apparent discrepancy in the energy consumption results: when the $L_2$-norm of the control input is used as a simple metric for energy consumption, the values are higher under event-triggered and self triggered control, whereas the detailed fuel consumption model shows lower values compared to time-driven control. This is due to the fact that $u_i^2$ penalizes CAVs when they decelerate, whereas this is not actually the case under a realistic fuel consumption model.

\begin{table*}\scriptsize
        \centering
        \begin{tabular}{|c|c|c|c|c|c|c|}
            \cline{1-7}
             &Item & \multicolumn{3}{|c|}{Event triggered} & Self Triggered &Time Triggered\\
            \cline{2-7}
            & Bounds & $s_v=0.5, s_x=1.5$ & $s_v=0.5, s_x=2$ & $s_v=0.5, s_x=2.5$ & $T_d=0.05 \ , T_{max}=0.5$  & $\Delta t = 0.05$\\
        \hline  
        \multirow{4}{*}{\makecell{$\alpha=0.1$ }} & Ave. Travel time & 19.61 & 19.73 & 19.65 & 19.65 & 19.42\\
        \cline{2-7}
        & Ave. $\frac{1}{2} u^2$ & 4.45 & 4.81  & 5.16  & 5.72  & 3.18\\
        \cline{2-7}
        & Ave. Fuel consumption &31.77 & 31.51 & 31.04 & 32.09 & 31.61\\
        \cline{2-7}
        &Computation load (Num of QPs solved) & 50\% (17853)&  47\% (16778) & 34\% (12168) & 13\% (4647) & 100\% (35443) \\
        \cline{2-7}
        & Num of infeasible cases   & 42  & 42 & 43 & 43& 315\\
        \hline
        \multirow{4}{*}{\makecell{$\alpha=0.25$ }}  & Ave. Travel time & 15.82 & 15.88 & 15.95 & 15.66 & 15.44\\
        \cline{2-7}
        & Ave. $\frac{1}{2} u^2$& 13.93 & 14.06 & 14.25 & 15.28 & 13.34\\
        \cline{2-7}
        & Ave. Fuel consumption & 52.12 & 51.69  & 51.42 & 54.03 & 55.81 \\
        \cline{2-7}
        &Computation load (Num of QPs solved) &51\% (14465) & 51\% (14403) & 48\% (13707) & 14\% (3954) & 100\% (28200)\\
        \cline{2-7}
        & Num of infeasible cases   & 27 & 27 & 28 & 27 & 341 \\
        \hline
                \multirow{4}{*}{\makecell{$\alpha=0.4$ }}  & Ave. Travel time & 15.4 &  15.46 & 15.53  & 15.24 & 15.01 \\
        \cline{2-7}
        & Ave. $\frac{1}{2} u^2$& 18.04 & 18.13 & 18.22 & 19.07 & 17.67\\
        \cline{2-7}
        & Ave. Fuel consumption & 53.155 & 52.77 & 52.42 & 54.89  & 56.5\\
        \cline{2-7}
        &Computation load (Num of QPs solved)    & 54\% (14089) & 53\% (14072) & 49\% (13573) & 15\% (4228)& 100\% (27412)\\
        \cline{2-7}
        & Num of infeasible cases   & 25 & 25 & 25 & 25  & 321 \\
        \hline       
                \multirow{4}{*}{\makecell{$\alpha=0.5$ }}  & Ave. Travel time & 15.05  & 15.11 & 15.17 & 14.88 & 14.63 \\
        \cline{2-7}
        & Ave. $\frac{1}{2} u^2$ &24.94 & 24.88 & 24.93 & 26.08 & 25.08\\
        \cline{2-7}
        & Ave. Fuel consumption & 53.65 & 53.41 & 53.21 & 54.89 & 56.93 \\
        \cline{2-7}
        &Computation load (Num of QPs solved) & 51\% (13764)  & 51\% (13758) & 50\% (13415) & 17\% (4743) & 100\% (26726) \\
        \cline{2-7}
        & Num of infeasible cases   & 20 & 20 & 20 & 20 & 341\\
        \hline        
        \end{tabular}
        \caption{CAV metrics under event-driven and time-driven control. }
        \label{Table I}
\end{table*}

\begin{figure}[H]
\centering
\includegraphics[scale=0.38]{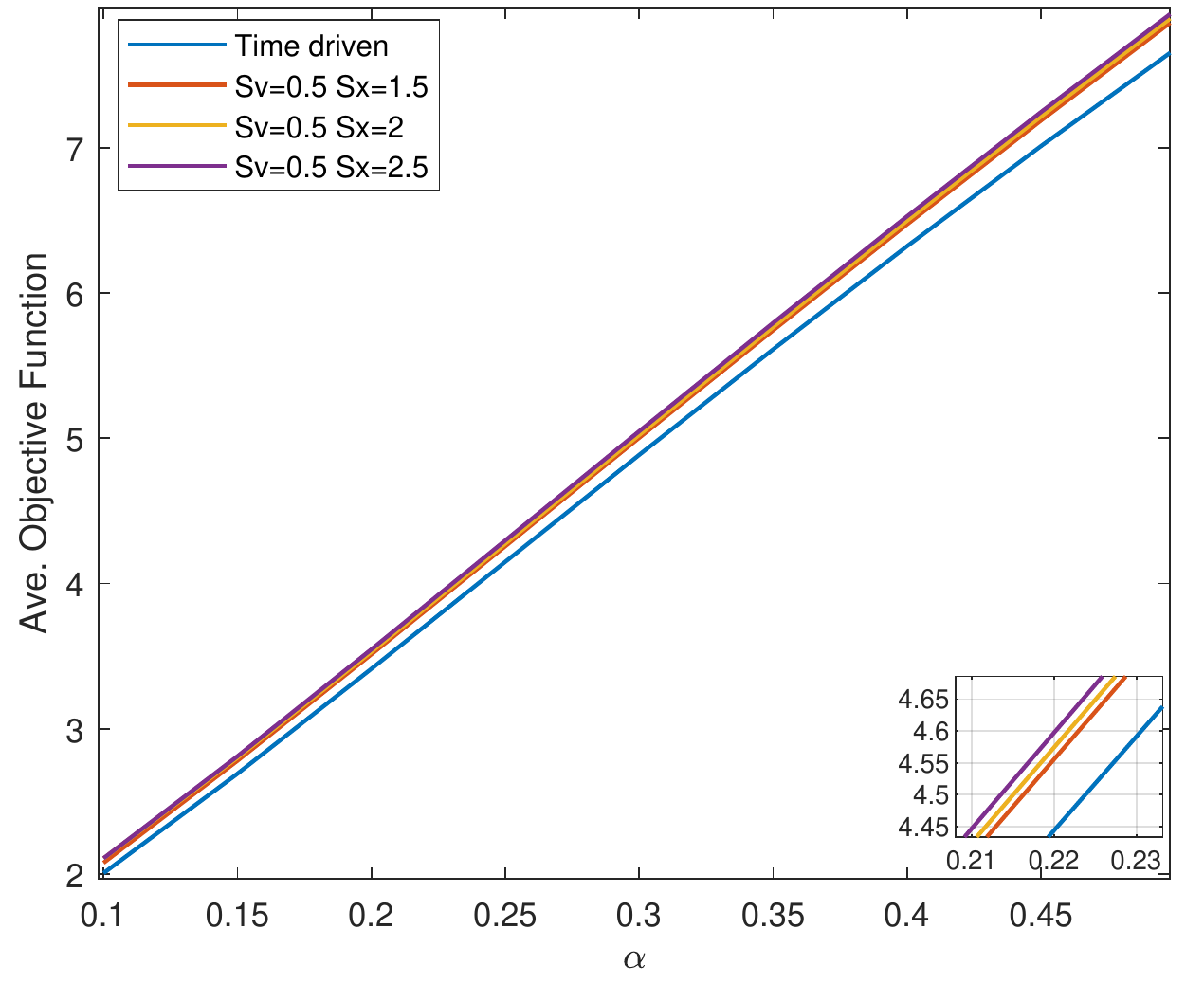} \caption{Average objective function value with respect to $\alpha$ for different selection of bounds.}
\label{fig:Objective function comparison}
\end{figure}

We can also visualize the results presented in the table by showing the variation of the average objective function in \eqref{eqn:energyobja_m} with respect to $\alpha$ for different choices of $[s_x,s_v]$. As seen in Fig. \ref{fig:Objective function comparison}, by selecting higher bounds (being more conservative) the objective function will also attain higher values, while the lowest cost (better performance) is reached under time-driven control.

\begin{figure}[t]
\centering
\includegraphics[scale=0.43]{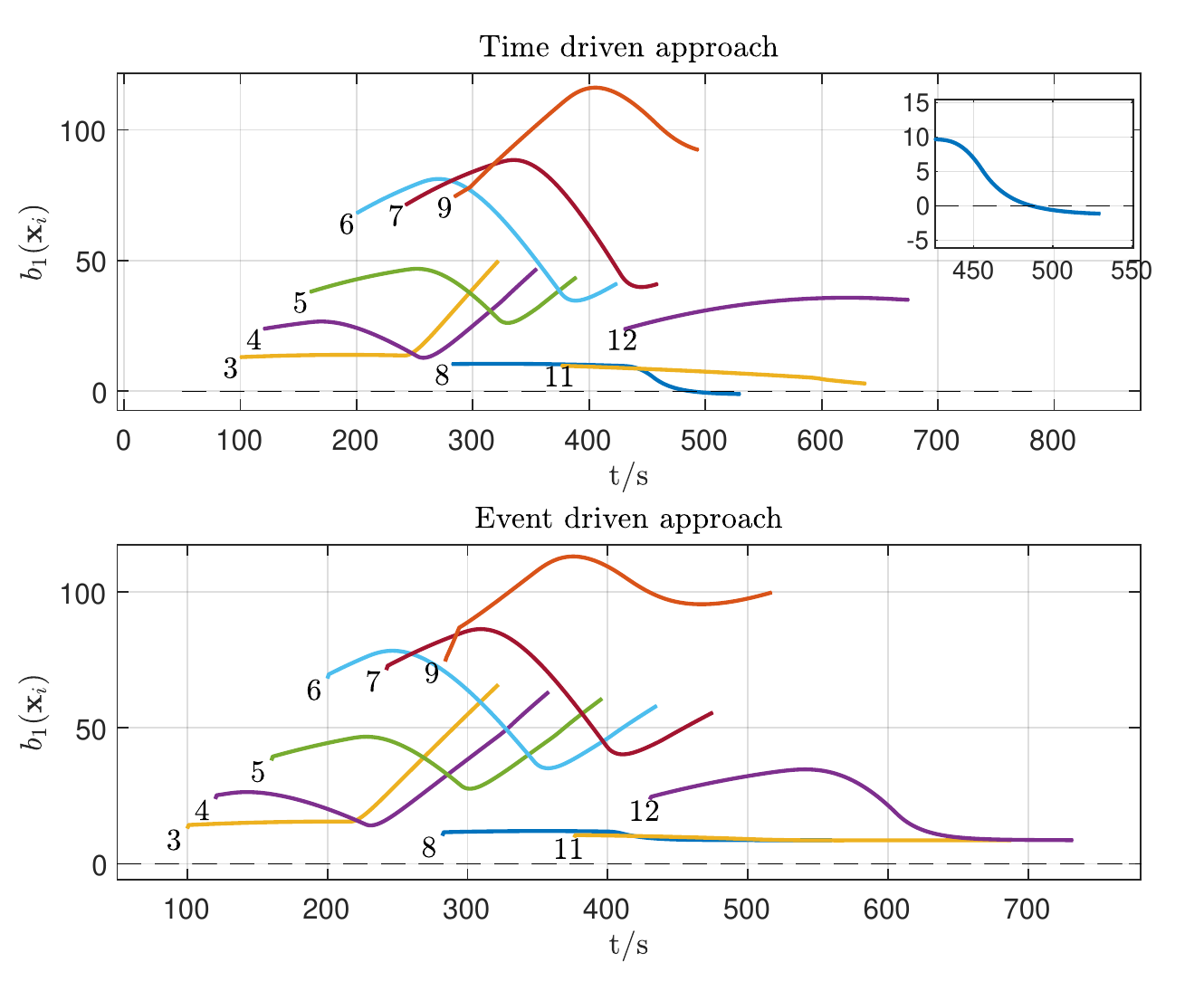} \caption{The variation of function $b_1(\textbf{x})$ for the time-driven and event-driven approaches.}
\label{fig:rear_end}
\end{figure}

{\bf Constraint violation}.
It is worth noting that an ``infeasible'' QP does not necessarily imply a constraint violation, since violating a CBF constraint does not always imply the violation of an original constraint in \eqref{Safety}, \eqref{SafeMerging}, and \eqref{VehicleConstraints1}. This is due to the conservative nature of a CBF whose intent is to 
\emph{guarantee} the satisfaction of our original constraints. 
In order to explicitly show how an infeasible case may lead to a constraint violation and how this can be alleviated by the event-triggering scheme, we simulated 12 CAVs in the merging framework of Fig. \ref{fig:merging} with the exact same parameter setting as before and with $S=[0.5,1.5]$ and $\beta = 5$. Figure \ref{fig:rear_end} shows the values of the rear-end safety constraint over time. One can see that the satisfaction of safety constraints is always guaranteed with the event-driven approach as there is no infeasible case and the value of $b_1(\textbf{x}(t))$ is well above the zero line. In contrast, we see a clear violation of the constraint in the time-driven scheme in the cases of CAVs 8 depicted by the blue line.

\begin{figure}[t]
\centering
\includegraphics[scale=0.4]{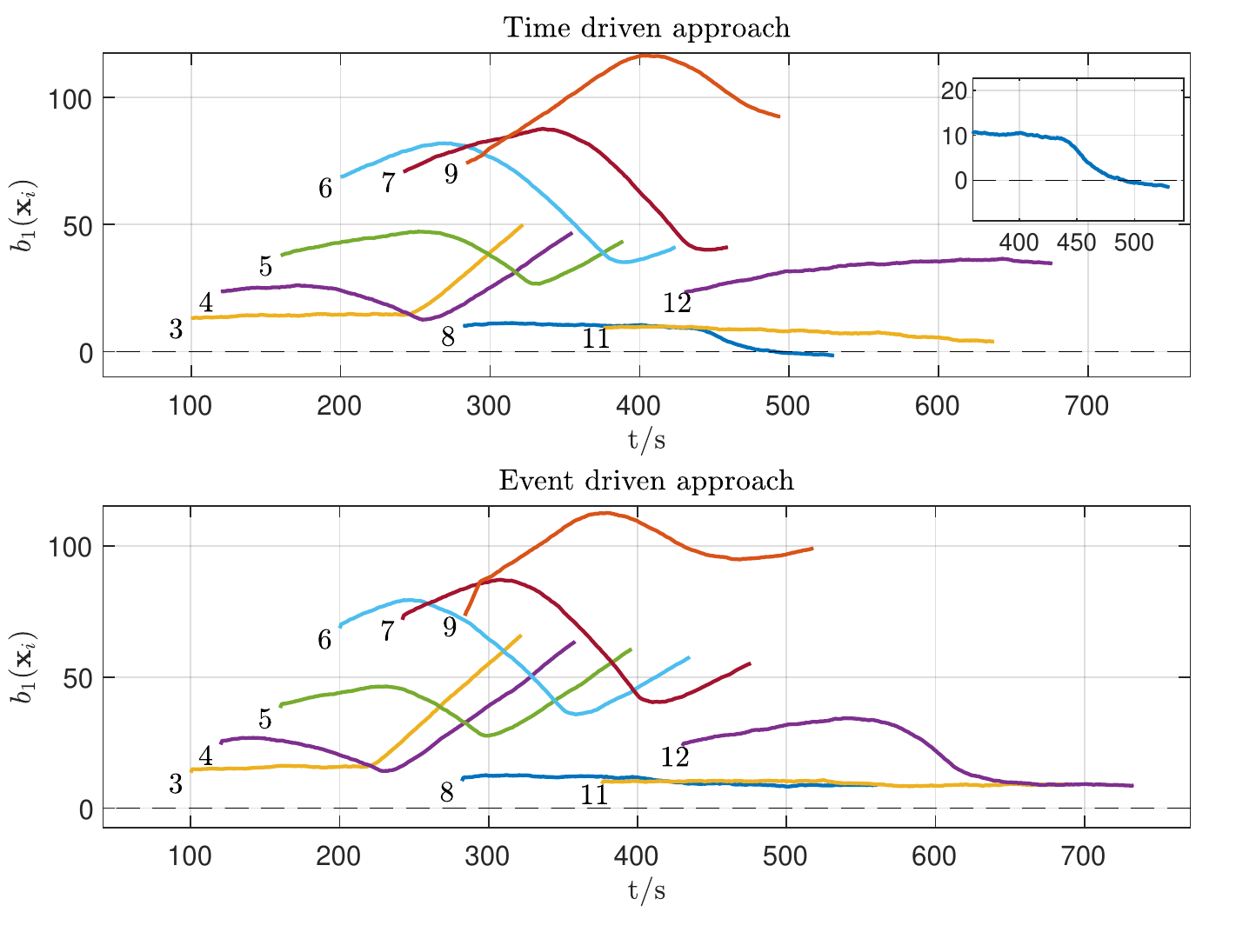} \caption{The variation of function $b_1(\textbf{x})$ for the time-driven and event-driven approaches in the presence of noise.}%
\label{fig:rear_end_noisy}
\end{figure}

\begin{figure}[t]
\centering
\includegraphics[scale=0.38]{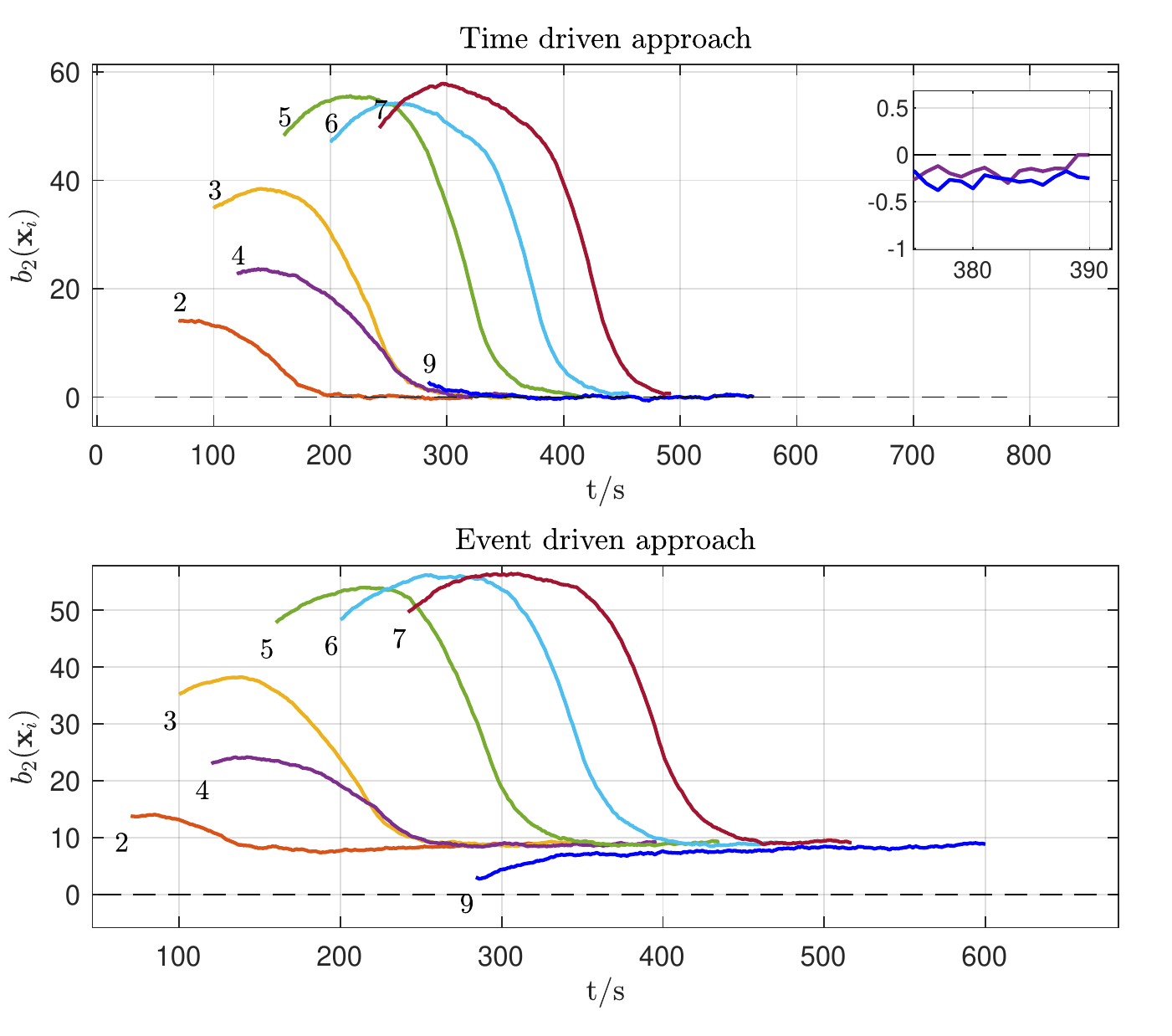} \caption{The variation of function $b_2(\textbf{x})$ for the time-driven and event-driven approaches in the presence of noise.}%
\label{fig:Lateral_noisy}
\end{figure}

{\bf Robustness}.
We have investigated the robustness of the event-driven scheme with respect to different forms of uncertainty, such as modeling and computational errors, by adding two noise terms to the vehicle dynamics: $\dot{x}_{i}(t) = v_{i}(t)+w_1(t)$, $\dot{v}_{i}(t) = u_{i}(t)+w_2(t)$,
where $w_1(t),w_2(t)$ denote two random processes defined in an appropriate probability space which, in our simulation, are set to be uniformly distributed over $[-2,2]$ and $[-0.2,0.2]$, respectively. We repeated the prior simulation experiment with added noise and results shown in Figs. \ref{fig:rear_end_noisy} and \ref{fig:Lateral_noisy}. We can see that the event-triggered scheme keeps the functions well away from the unsafe region (below $0$)
in contrast to the time-driven approach where we observe constraint violations due to noise, e.g., CAV 8 in Fig. \ref{fig:rear_end_noisy} and CAVs 4 and 9 in Fig. \ref{fig:Lateral_noisy}.

\section{CONCLUSIONS}
The problem of controlling CAVs in conflict areas of a traffic network subject to hard safety constraints can be solved through a combination of tractable optimal control problems and the use of CBFs. These solutions can be derived by discretizing time and solving a sequence of QPs. However, the feasibility of each QP cannot be guaranteed over every time step. When this is due to the lack of a sufficiently high control update rate, we have shown that this problem can be alleviated through an event-triggering scheme, while at the same time reducing the need for communication among CAVs, thus lowering computational costs and the chance of security threats. Simulation results illustrate how safety constraint violations can be prevented using this event-triggering scheme. Ongoing work is targeted at eliminating all possible infeasibilities through the use of sufficient conditions based on the work in \cite{XIAO2022inf} added to the QPs, leading to complete solutions of CAV control problems with full safety constraint guarantees.

\label{sec:conclude}





\bibliographystyle{unsrt}

\begin{thebibliography}{10}

\bibitem{li2013survey}
D.~Wen L.~Li and D.~Yao.
\newblock A survey of traffic control with vehicular communications.
\newblock {\em IEEE Trans. on Intelligent Transportation Systems}, 15(1):pp.
  425--432, 2013.

\bibitem{f60c6852682d4629b75f458c50adbbf1}
D.~{de Waard}, C.~Dijksterhuis, and KA. Brookhuis.
\newblock Merging into heavy motorway traffic by young and elderly drivers.
\newblock {\em Accident Analysis \& Prevention}, 41(3):pp. 588--597, 2009.

\bibitem{Schrank20152015UM}
T.~Lomax D.~Schrank, B.~Eisele and J.~Bak.
\newblock 2015 urban mobility scorecard.
\newblock 2015.

\bibitem{7562449}
J.~Rios-Torres and A.~A. Malikopoulos.
\newblock A survey on the coordination of connected and automated vehicles at
  intersections and merging at highway on-ramps.
\newblock {\em IEEE Trans. on Intelligent Transportation Systems}, 18(5):pp.
  1066--1077, 2017.

\bibitem{xu2019grouping}
H.~Xu, S.~Feng, Y.~Zhang, and L.~Li.
\newblock A grouping-based cooperative driving strategy for cavs merging
  problems.
\newblock {\em IEEE Trans. on Vehicular Technology}, 68(6):pp. 6125--6136,
  2019.

\bibitem{7313484}
J.~Rios-Torres, A.~Malikopoulos, and P.~Pisu.
\newblock Online optimal control of connected vehicles for efficient traffic
  flow at merging roads.
\newblock In {\em 2015 IEEE 18th International Conf. on Intelligent
  Transportation Systems}, pages 2432--2437. IEEE, 2015.

\bibitem{XIAO2021109333}
W.~Xiao and C.G. Cassandras.
\newblock Decentralized optimal merging control for connected and automated
  vehicles with safety constraint guarantees.
\newblock {\em Automatica}, 123:109333, 2021.

\bibitem{Zhang2018}
Y.F Zhang and C.G. Cassandras.
\newblock Decentralized optimal control of connected automated vehicles at
  signal-free intersections including comfort-constrained turns and safety
  guarantees.
\newblock {\em Automatica}, 109:p. 108563, 11 2019.

\bibitem{cao2015cooperative}
W.~Cao, M.~Mukai, T.~Kawabe, H.~Nishira, and N.~Fujiki.
\newblock Cooperative vehicle path generation during merging using model
  predictive control with real-time optimization.
\newblock {\em Control Engineering Practice}, 34:98--105, 2015.

\bibitem{mukai2017model}
M.~Mukai, H.~Natori, and M.~Fujita.
\newblock Model predictive control with a mixed integer programming for merging
  path generation on motor way.
\newblock In {\em 2017 IEEE Conf. on Control Technology and Applications
  (CCTA)}, pages 2214--2219. IEEE, 2017.

\bibitem{Xiao2019}
W.~Xiao and C.~Belta.
\newblock Control barrier functions for systems with high relative degree.
\newblock In {\em Proc. of 58th IEEE Conf. on Decision and Control}, pages
  474--479, Nice, France, 2019.

\bibitem{XIAO2021109592}
W.~Xiao, C.G. Cassandras, and C.~Belta.
\newblock Bridging the gap between optimal trajectory planning and
  safety-critical control with applications to autonomous vehicles.
\newblock {\em Automatica}, 129:109592, 2021.

\bibitem{CBF_QP(2017)}
A.~Ames, X.~Xu, J.W. Grizzle, and P.~Tabuada.
\newblock Control barrier function based quadratic programs for safety critical
  systems.
\newblock {\em IEEE Trans. on Automatic Control}, 62(8):3861--3876, 2017.

\bibitem{ong2018event}
P.~Ong and J.~Cort{\'e}s.
\newblock Event-triggered control design with performance barrier.
\newblock In {\em 2018 IEEE Conf. on Decision and Control (CDC)}, pages
  951--956. IEEE, 2018.

\bibitem{taylor2020safety}
Taylor~A. J., Ong P., Cortés J., and Ames~A. D.
\newblock Safety-critical event triggered control via input-to-state safe
  barrier functions.
\newblock {\em IEEE Control Systems Letters}, 5(3):749--754, 2020.

\bibitem{Xiao2021EventTriggeredSC}
W.~Xiao, C.~Belta, and C.~G. Cassandras.
\newblock Event-triggered safety-critical control for systems with unknown
  dynamics.
\newblock In {\em 2021 60th IEEE Conf. on Decision and Control (CDC)}, pages
  540--545, 2021.

\bibitem{XIAO2022inf}
W.~Xiao, C.~A. Belta, and C.~G. Cassandras.
\newblock Sufficient conditions for feasibility of optimal control problems
  using control barrier functions.
\newblock {\em Automatica}, 135:109960, 2022.

\bibitem{xu2020general}
H.~Xu, W.~Xiao, C.~G. Cassandras, Y.~Zhang, and L.~Li.
\newblock A general framework for decentralized safe optimal control of
  connected and automated vehicles in multi-lane signal-free intersections.
\newblock {\em IEEE Trans. on Intelligent Transportation Systems}, pages 1--15,
  2022.

\bibitem{9682916}
W.~Xiao and C.G. Cassandras.
\newblock Decentralized optimal merging control for connected and automated
  vehicles on curved roads.
\newblock In {\em 2021 60th IEEE Conf. on Decision and Control (CDC)}, pages
  2677--2682, 2021.

\bibitem{Vogel2003}
K.~Vogel.
\newblock A comparison of headway and time to collision as safety indicators.
\newblock {\em Accident Analysis \& Prevention}, 35(3):427--433, 2003.

\bibitem{kamal2012model}
M.~A.~S. Kamal, M.~Mukai, J.~Murata, and T.~Kawabe.
\newblock Model predictive control of vehicles on urban roads for improved fuel
  economy.
\newblock {\em IEEE Trans. on Control Systems Technology}, 21(3):831--841,
  2012.

\end{thebibliography}

\end{document}